\documentclass[lettersize,journal]{IEEEtran}

\usepackage{array}
\usepackage[caption=false,font=normalsize,labelfont=sf,textfont=sf]{subfig}
\usepackage{textcomp}
\usepackage{stfloats}
\usepackage{url}
\usepackage{verbatim}
\usepackage{graphicx}
\usepackage{subfig}
\usepackage{cite}
\usepackage{dirtytalk}
\usepackage{changepage}
\usepackage{algpseudocode}
\usepackage{algorithm}
\usepackage{amsmath,amsfonts}
\usepackage[dvipsnames]{xcolor}
\usepackage{scalerel}
\usepackage{tikz}
\usetikzlibrary{svg.path}

\definecolor{orcidlogocol}{HTML}{A6CE39}
\tikzset{
  orcidlogo/.pic={
    \fill[orcidlogocol] svg{M256,128c0,70.7-57.3,128-128,128C57.3,256,0,198.7,0,128C0,57.3,57.3,0,128,0C198.7,0,256,57.3,256,128z};
    \fill[white] svg{M86.3,186.2H70.9V79.1h15.4v48.4V186.2z}
                 svg{M108.9,79.1h41.6c39.6,0,57,28.3,57,53.6c0,27.5-21.5,53.6-56.8,53.6h-41.8V79.1z M124.3,172.4h24.5c34.9,0,42.9-26.5,42.9-39.7c0-21.5-13.7-39.7-43.7-39.7h-23.7V172.4z}
                 svg{M88.7,56.8c0,5.5-4.5,10.1-10.1,10.1c-5.6,0-10.1-4.6-10.1-10.1c0-5.6,4.5-10.1,10.1-10.1C84.2,46.7,88.7,51.3,88.7,56.8z};
  }
}

\newcommand\orcidicon[1]{\href{https://orcid.org/#1}{\mbox{\scalerel*{
\begin{tikzpicture}[yscale=-1,transform shape]
\pic{orcidlogo};
\end{tikzpicture}
}{|}}}}

\usepackage[colorlinks=true,linkcolor=black,anchorcolor=black,citecolor=black,filecolor=black,menucolor=black,runcolor=black,urlcolor=black]{hyperref} 

\begin{document}

\title{IoT-based Android Malware Detection Using Graph Neural Network With Adversarial Defense}

\author{Rahul Yumlembam$^{\textsuperscript{\orcidicon{0000-0002-0313-5731}}}$, Biju Issac$^{\textsuperscript{\orcidicon{0000-0002-1109-8715}}}$,~\IEEEmembership{Senior Member,~IEEE}, Seibu Mary Jacob$^{\textsuperscript{\orcidicon{0000-0001-9033-794X
}}}$ and Longzhi Yang$^{\textsuperscript{\orcidicon{0000-0003-2115-4909
}}}$
\thanks{Rahul Yumlembam, Biju Issac, and Longzhi Yang are with the Department of Computer and Information Sciences, Northumbria University, Newcastle, UK (email: rahul.yumlembam@northumbria.ac.uk, bissac@ieee.org, longzhi.yang@northumbria.ac.uk). Corresponding author: Biju Issac}
\thanks{Seibu Mary Jacob is with the School of Computing, Engineering \& Digital Technologies, Teesside University, Middlesbrough, UK (email: s.jacob@tees.ac.uk).}
}

\markboth{IEEE Internet Of Things Journal}%
{Shell \MakeLowercase{\textit{et al.}}: A Sample Article Using IEEEtran.cls for IEEE Journals}

\IEEEpubid{\begin{minipage}{\textwidth}\ \\[12pt] \centering
  Copyright (c) 2022 IEEE. Personal use of this material is permitted. However, permission to use this material for any other purposes must be obtained from the IEEE by sending a request to pubs-permissions@ieee.org.
\end{minipage}}


\maketitle

\begin{abstract}
Since the Internet of Things (IoT) is widely adopted using Android applications, detecting malicious Android apps is essential. In recent years, Android graph-based deep learning research has proposed many approaches to extract relationships from the application as a graph to generate graph embeddings. First, we demonstrate the effectiveness of graph-based classification using Graph Neural Networks (GNN) based classifier to generate API graph embedding. The graph embedding is used with ‘Permission’ and ‘Intent’ to train multiple machine learning and deep learning algorithms to detect Android malware. The classification achieved an accuracy of 98.33\% in CICMaldroid and 98.68\% in Drebin dataset. However, the graph-based deep learning is vulnerable as an attacker can add fake relationships to avoid detection by the classifier. Second, we propose a Generative Adversarial Network (GAN) based algorithm named VGAE-MalGAN to attack the graph-based GNN Android malware classifier. The VGAE-MalGAN generator generates adversarial malware API graphs, and the VGAE-MalGAN substitute detector (SD) tries to fit the detector. Experimental analysis shows that VGAE-MalGAN can effectively reduce the detection rate of GNN malware classifiers. Although the model fails to detect adversarial malware, experimental analysis shows that retraining the model with generated adversarial samples helps to combat adversarial attacks. 
\end{abstract}

\begin{IEEEkeywords}
Internet of Things, Graph Neural Network, Generative Adverserial Network, Android, Machine Learning, Deep Learning   
\end{IEEEkeywords}

\section{Introduction}\label{Intro}

\begin{figure}[!hbt]
\centering
\includegraphics[width=5.5cm]{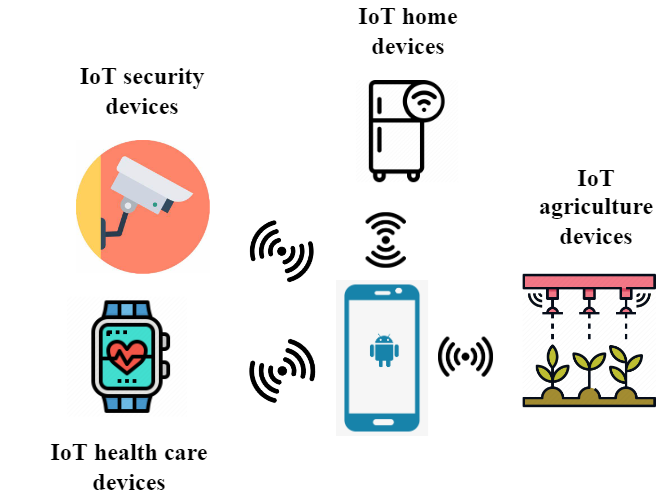}
\caption{\small{The different IoT devices connected to Android platform}}
\label{fig:Android_IoT}
\end{figure}

In the recent years there has been an increase in the usage of IoT devices to improve the quality of our lives. IoT device utility can range from smart homes to industrial automation. These devices must interact with the user for data exchange or communication. One of the most common ways to control these IoT devices is through applications installed on a smartphone. Through these applications, the users can communicate with various IoT devices, say, through monitoring the the room's temperature, live video feed, heart rate, the water level in agricultural settings, etc., as shown in Fig. \ref{fig:Android_IoT}. The applications used to control this device holds critical and valuable information, which is very lucrative for attackers. For example, an attacker who has access to the application can monitor the CCTV camera or access the health information stored on a smartphone. Malware on IoT applications can significantly violate the privacy of any IoT user. 
\IEEEpubidadjcol
Malicious Android applications can therefore act as a gateway to attack IoT devices. These malicious IoT based Android apps can get installed accidentally through user lapses in judgment or from the apps installed from unknown sources. According to Nokia Threat Intelligence Report 2020, Android \cite{Nokia} accounts for 26.64\% of infections across all platforms and IoT devices are now responsible for 32.72\% of all  infections observed in mobile networks, up from  16.17\% in the previous year.

 Android malware is a malicious application that steals sensitive information, violates user privacy, or performs any action the user did not authorize. According to AV-test \cite{avtest}, in 2021, 3.39 million malware emerged in the market. It is crucial to identify applications that can harm users. In 2021, Kaspersky Android mobile products and technologies detected 3,464,756 malicious installation packages, 97,661 new mobile banking trojans and 17,372 new mobile ransomware trojans \cite{Shishkova}.

 There are two kinds of malware detection analysis: static analysis and dynamic analysis. In the static analysis, the static features of the application, such as permission, intents, signature etc., are analyzed. In dynamic analysis, the dynamic features of the application, such as network flow information, app actions sequence etc., are analyzed. 

We have opted for static analysis, since every possible branch of the code must execute for effective dynamic feature generation. With the rapid pace of malware generation, the development of different techniques for identifying and analyzing them is a critical requirement.
 
In recent years, Android graph-based deep learning research has proposed many approaches to extract relationships from the application as a graph to generate graph embedding. For example, Hindroid \cite{hou2017hindroid} extracts API relationships based on Code block and API-Invoke method, whereas MalScan \cite{wu2019malscan} extracts function call graph. Similarly, in DroidMiner \cite{yang2014droidminer}, a component dependency graph and a component behaviour graph are constructed. The extracted graphs need to exist in a format suitable for the downstream task. To this end, recent papers have proposed to use GNN in \cite{feng2020android} and \cite{cai2021learning}. In this work, we first demonstrate the effectiveness of the graph-based technique by using API graph embedding along with Permission and Intent as features for classification. The API graph is generated based on code block id\cite{hou2016droiddelver}\cite{hou2017hindroid}. The edges between the nodes (API) in the API graph represent relationships between different APIs. Centrality measures can express this relationship by measuring a node's importance relative to all the other nodes in the graph. Different centrality measures are extracted from the API graph to train a GNN \cite{scarselli2008graph},\cite{hamilton2017inductive} to generate graph embedding of each Android Application. The generated graph embedding is combined with Permission, Intent and used to train multiple machine learning and deep learning algorithms. The trained GNN malware classifier acts as a model where no gradient information is accessible. We then propose an adversarial architecture named VGAE-MalGAN to attack graph-based Android malware classifier.

Although recent works to extract graph embedding from relationship graphs have proven resilient against malware attacks, there has been little study on how dummy relationship contamination can fool an Android malware classifier. The recent work in \cite{wan2021adversarial} proposed an algorithm called Grabnel to attack GNN model. Although this algorithm can successfully attack the model, it has no mechanism to preserve the original semantics of the malware API graph as in our attack. We aim to address the mentioned research gap in the work done.

 The main contributions of our research are as follows:
\begin{itemize}
    \item We demonstrate the effectiveness of GNN in generating graph embedding using centrality features of an API graph used along with ‘Permission’ and ‘Intent’ to improve malware classification.
     \item We propose a new approach named VGAE-MalGAN that can effectively  add  nodes  and  edges  to  an  existing  API  graph and dynamically generate an adversarial Android malware API graph which can fool the GNN based malware classifier trained using GNN. It will still preserve the original semantics of the malware API graph. VGAE-MalGAN comprises of a Generator and substitute detector. The Generator is a modified version of Variation Graph Auto Encoder, and the substitute detector is a GraphSAGE model.
    \item We demonstrate that the model can be hardened against attack by VGAE-MalGAN through retraining and can restore high malware detection accuracy.
    
\end{itemize}

The paper is organised as follows. Section \ref{Overview} is the overview of the work done, section \ref{preliminary_study_1} is the preliminary study, section \ref{proposed_method} is the proposed method, section \ref{evaluation} is evaluation, section \ref{related_works} is related works, and section \ref{conclusion} is conclusion.

\section{Overview of the work}\label{Overview}
\begin{figure}
\centering
\includegraphics[width=9cm]{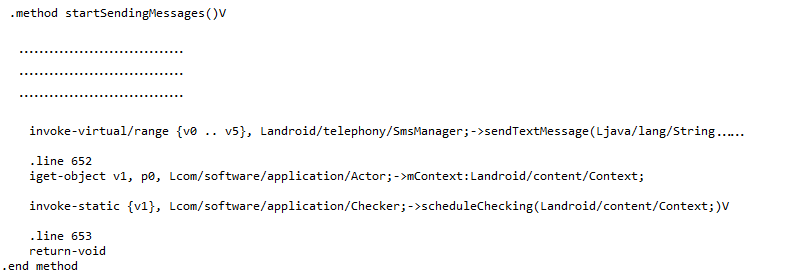}
\caption{\small{Sample Smali code of an Android application}} 
\label{smali_code}
\end{figure}
Chatzoglou et al. in their recent work \cite{Chat2022} closely examined more than forty top Android official apps belonging to six diverse mainstream categories of IoT devices and found that majority of IoT based Android apps remain susceptible to a range of security and privacy issues. Java source code written for the Android application compiles into .class files by Java compiler and Android SDK (Software Development Kit) converts the .class files into .dex files, also known as Dalvik executables. The Android assets packaging tool packages .dex files and all the resources (images, video files, audio files, XML files, etc.) into an Android application package (APK). APKs are then used for distribution and installation by smartphones running the Android operating system. The .dex files are not in a human-readable format. In order to extract information from the .dex files, Apk tool converts them into Smali files, a human-readable version of the .dex file. Smali code is a representation of the app code using the Android Dalvik opcodes. Fig. \ref{smali_code} shows a sample Smali code. Another important file used in this work is called the Android manifest files. The manifest file describes essential information about an application such as code namespace, components of the application, permission it needs to access part of the system or other application, hardware, and software features that the app requires. 

To classify an Android application into Malware or Benign, we follow the following steps:
(1) Decompile: Using Apk-tool\cite{Apk-Tool} we extract smali files and manifest files of each application in the dataset.
(2) Feature Extraction: From the extracted smali files, the APIs from each application, along with their code block ids, are extracted. Extracted APIs are then assigned a unique global id. This unique id is similar across all the applications. Using the Linear regression feature selection APIs are selected based on feature importance weight. Local graphs and a global graph are constructed based on selected APIs. Local graphs are individual graphs extracted from each application, whereas the local graphs combine to form a global graph. Centrality features are extracted from the global graph to train a GNN. The trained GNN generates graph embedding of each application. We then extract Permission and Intent features from the Manifest file.
(3) Malware Detection: Generated graph embedding is used along with Permission and Intents as features to train machine learning algorithms that classify the application into benign or malware. Fig. \ref{Overall_archi} shows the overall architecture of Malware Detector.

VGAE-MalGAN shown in Fig. \ref{VGAEMalGAN} is used to fool the trained classifier by generating adversarial API graphs. VGAE-MalGAN is a  VGAE \cite{kipf2016variational} based GAN \cite{goodfellow2014generative} which generates adversarial Malware API graphs that could fool  Malware detectors. The VGAE used in this work is modified from its initial proposal to accommodate sparsely connected graphs, typical in API graphs. The generated adversarial examples are augmented to the original dataset to retrain and increase their detection capabilities.

\section{Preliminary Study}\label{preliminary_study_1}
This section describes the background information, how the experiment represents Android applications, and how the Android applications are classified into benign and malware.

\subsection{Datasets Used}
The datasets used for this work are CICMalDroid 2020\cite{mahdavifar2020dynamic} and Drebin dataset\cite{arp2014drebin}.
CICMalDroid 2020 contains 17,341 samples, and as reported on the original dataset website, only  13,077 samples ran successfully without any errors.The malware type in CICMaldroid includes Adware, Banking Malware, SMS Malware, and Mobile Riskware. Our experiment successfully extracted features from 15,848 applications (3696 benign and 12,152 malware). This dataset also includes other static and dynamic features along with the APKs. Drebin dataset contains 5,560 malware APKs. For the benign APKs, only SHA-256 of the application are available on the website. Using Androzoo\cite{allix2016androzoo} we are able to download 50,901 benign applications listed in the Drebin benign SHA-256. The remaining benign SHA-256 returned invalid from the Androzoo API. The malware dataset in Drebin includes Backdoor, Adware, worm and Trojan. We performed a 70/30 split on each dataset for training and testing.

\subsection{Feature Extraction and Malware classification}
\subsubsection{API extraction and API selection}
From the training set of the datasets, APIs used in an application are extracted by parsing the smali files of an application. As an example in Fig. \ref{smali_code} from the smali code segment, \say{$\text{Landroid/telephony/SmsManager}\rightarrow \text{sendTextMessage}$}
along with \say{$\text{Lcom/software/application/Checker}\rightarrow\text{scheduleChecking}$} will be extracted as an API. Due to space constraints, we did not show the other APIs in the code segment. The extracted APIs are then assigned a unique global identifier. The unique global identifier is similar across all the applications.

The number of unique APIs extracted is enormous. For example, in the data set CICMaldroid2020, more than 1.7 million unique APIs are extracted. Using Sklearn feature selection method selectFromModel, we extract the top 7000 APIs based on features importance weight to reduce computation time and increase the efficiency. We select the API by training a Linear Regression using API count as features.

We found out that increasing the number of selected APIs does not increase classification accuracy through different experiments. After the feature selection, each application $A_{j}$ will have 7000 API where $A_{j}={\{a_{1}…a_{7000}\}}$.

\subsubsection{API graph Construction} Two kinds of graphs, i.e., local graph and global graph, are constructed to leverage the relationship among the APIs. To construct an API graph, we first define what is a code block using definition from\cite{hou2017hindroid}; a code block is the code segment between a $.method$ and $.endmethod$ in smali files. To construct the local API graph, we construct an adjacency matrix $A$ for each application where $A_{i,j} = 1$, if API $i$ and $j$ belong to the same code block; else it is set to 0. As an example, in Fig. \ref{smali_code}  
\say{$\text{Landroid/telephony/SmsManager} \rightarrow \text{sendTextMessage}$} and \say{$\text{Lcom/software/application/Checker}\rightarrow \text{scheduleChecking}$} are nodes in the graph with an edge between them. After extracting the adjacency matrix of each application, we construct the Global graph $O$ where $O_{i,j}=1$ if in any application in the training set, API $i$ and $j$ co-occur in the same
code block; else it is set to 0. The overall API graph contains APIs only from the training set to mimic real-world scenarios. It is worthy to note that API that belongs to different code blocks also forms a relationship through intermediate API. As an example, API $i$ and $j$ belong to a code block, and API $k$ and $j$ belong to a different code block. There exists a relationship between API $i$ and $k$ through intermediate API $j$.

\subsubsection{Centrality Feature Extraction} 
One popular way of characterizing the role of a node in a network is by using one or more centrality measures. These measures aim to quantify the capacity of a node to influence or be influenced by other nodes. From the global graph, we extract five different centrality measures of each node, namely Degree centrality, Betweenness centrality, Closeness centrality, EigenVector centrality and Page rank. The five centrality features acts as features of each node.

\subsubsection{Importance of API graph and Centrality features} 
APIs reveal exciting information about an application, and the relationship between the APIs will be different for benign and malware. The API graph effectively represents information where the edge between the nodes(API) represents the relationship between the nodes(API).
We analyse one SMS malware application
{\scriptsize SHA256-000e7149ab7550ef605c2b22cb1beaffbee9219699661d89158d490a3ffa393a} to help demonstrate why API graph reveals useful information about an application. From Fig. \ref{smali_code}, it is seen that \say{$\text{Landroid/telephony/SmsManager}\rightarrow \text{sendTextMessage}$} which is used to send text message is first called and an application-specific API defined by the developer, \say{$\text{Lcom/software/application/Checker}\rightarrow \text{scheduleChecking}$} is then called. This sequential call of APIs forms a relationship between the two APIs. When the Checker.class is analyzed, an alarm event sends a text message to a  number every 30 seconds using a broadcast receiver created. It uses \say{$\text{Landroid/app/PendingIntent}\rightarrow \text{getBroadcast}$} to listen to an alarm event. When it receives the broadcast, it sends a text message which then sets a new alarm  30 seconds from the current time using \say{$\text{/app/AlarmManager}\rightarrow \text{set}$}. \say{$\text{/app/AlarmManager}\rightarrow \text{set}$} which is an alarm manager API to make an application run at some point of time in future. The connection among the APIs reveals useful information captured in the API graph.

\begin{figure}[!hbt]
\centering
\includegraphics[width=8.5cm]{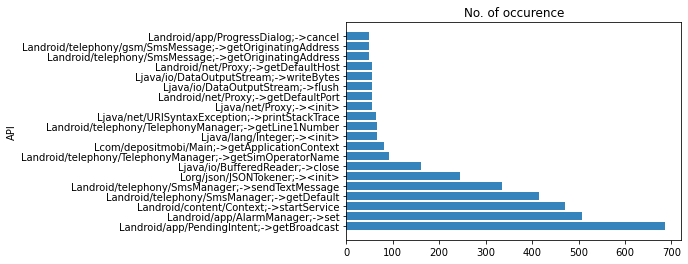}
\caption{\small{Most common central APIs in SMS malware}} 
\label{sms_apis}
\end{figure}

To determine the importance of each node in the application, we calculate different centrality measures for each node in the application. We took a sample of 100 SMS malware applications and 100 benign applications. Let “S” be the set of the top 5\% nodes most central to a sample of SMS malware applications, and let “B” be the top 5\% nodes most central to a sample of benign applications. We took the difference of the two sets, i.e., all elements in S that are not in B. Fig. \ref{sms_apis} shows the 20 most common APIs in the difference between the two sets. From the figure, \say{$\text{Lcom/depositmobi/Main}\rightarrow     \text{getApplicationContext}$} is among the top APIS used in SMS malware. Lcom/depositmobi is a well-known SMS malware. A lot of the applications repackage previously existing malware applications. From all the above analysis, we can safely infer that each application  API graph and centrality measure are powerful features that can describe an application.

\subsubsection{Importance of Permission and Intents:}
 According to Android SDK (Software Development Kit), any functionality the application uses declares itself inside the manifest file. Thus, Permission and Intent can summarize application functionality. Malware application needs access to more sensitive Permission and Intent and, thus, they become an important differentiator between a Malware application and a benign application. We extract the permission and Intent from the manifest file. If the application uses corresponding Permission or Intent, the corresponding position of Permission or Intent sets to 1; otherwise, 0 in the feature vector.

\subsubsection{Graph Embedding Generation}
Using Graph Neural Networks (GNN), we generate Android application API graph embedding such that similar graphs are embedded close together. The cosine similarity between the vector describing the graphs is high if the graphs are similar.

GNN maps each node in a graph to a d-dimensional embedding such that similar nodes in the graph are embedded close together. More formally, GNN can be defined as given a Graph $G(V,E)$ where $V$ is a finite set of nodes $v$ and edges $E\subseteq \left \{ (u,v)\subseteq V \right \}$. Neighborhood of node $v$ is denoted by $N(v)$. In each layer $t>0$, we compute a new representation for node $v$ using :
\begin{equation}
f^{(t)}(v)=\sigma (f^{(t-1)}(v).W_{1}^{(t)}+\sum_{w\epsilon N(v)}^{}f^{(t-1)}(w).W_{2}^{(t)})
\end{equation}

where, $W_{1}$ and $W_{2}$ are trainable parameter matrices and $\sigma$ denotes the element-wise non-linearity (e.g. a tanh or ReLu). We trained two GNN variants to generate the embedding, i.e., GCN (Graph Convolutional Network) and GraphSAGE. GCN is a variation of GNN where the main idea is to transform information from the neighbors and combine them. It generates node embedding based on local network neighborhoods. In GCN, $f^{(t-1)}(w)$ is given by equation 2, which performs the average of previous layer embeddings. In each layer $t>0$, we compute a new representation for node w using:

\begin{equation}
f^{(t-1)}(w)=\frac{f_{w}^{(t-1)}}{\left\|N(v)\right\|},
\end{equation}

\begin{figure} [!hbt]
\centering
\includegraphics[width=8cm]{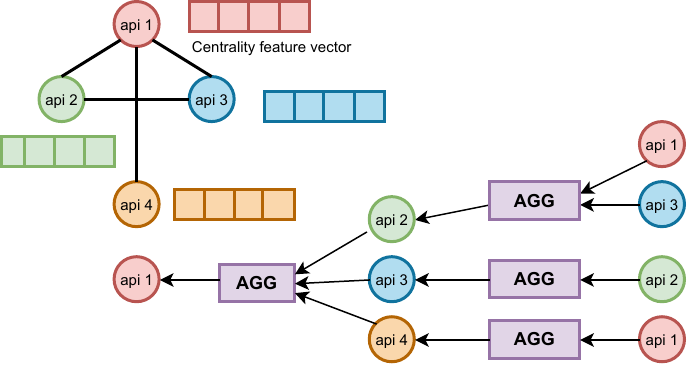}
\caption{\small{Sample GNN Computational Graph}} 
\label{GNN computation graph}
\end{figure}

GraphSAGE (GS) is a variation of GCN where the sum defined over the neighbourhood is replaced by generalised aggregation function which is permutation invariant differentiable function and the outer sum is replaced by concatenation. In each layer $t>0$, we compute a new representation for the node $v$ using:.
\begin{equation}
f^{(t)}(v)=\sigma (f_{merge}(f^{(t-1)}(v).W_{1},f_{W_{2}}^{aggr}( f^{(t-1)}(w)\| w\epsilon N(v))))
\end{equation}
where, $f_{W_{2}}^{aggr}$ is the generalised aggregation function. The generalised aggregation function are parameterized by trainable parameter matrices $W_2$. The generalised aggregation function generates neighbor embedding. Previous layer embedding of node $v$ given by  $f^{(t-1)}(v)$ parametirized by $W_1$ is then concatenated with $f_{W_{2}}^{aggr}$ using $f_{merge}$. In this work we use mean aggregation function. Fig. \ref{GNN computation graph} shows the computational graph of an example node API 1. The computational graph is a two-layer GNN where we calculated aggregation $AGG$  using equation 1 for GCN and equation 2 for GraphSAGE.
\paragraph{Graph Embedding} The embedding of a graph can be computed using:
\begin{equation}
f_{GNN}(G)=\sum_{v\epsilon V(G)}f^{(T)}(v),
\end{equation}
where $T>0$ denotes the last layer, $V(G)$ denotes all the nodes in the graph.
The global centrality measures act as features to train the GCN and GraphSAGE network. Both the GNN networks have a similar architecture consisting of three layers of size $(5 \times 32)$, $(32 \times 32)$, and $(32 \times 32)$, followed by a global mean pool layer which converts the output into $1 \times 32$  vector, which is the graph embedding of the application. The network trains in a supervised way. The embedding generated by the global mean pool layer is used as input to a dropout layer with a dropout rate of 0.5, followed by a fully connected layer that outputs the probability of an application being benign or malware. We use the Adam optimization algorithm for optimizing the network parameters with the learning rate set to 0.001, beta1 set to 0.9, beta2  set to 0.999, and epsilon set to 1e-08. The network's loss is calculated using the cross-entropy function and mean aggregation as the aggregator for GraphSAGE during training. From the result depicted in Table \ref{overall performance}, it is clear that GraphSAGE provides better classification accuracy; therefore, to generate the embedding, we pass the API graph of each application to the GraphSAGE network generating an embedding $\vec{emb}$ 
of size 1 $\times$ 32 for each application. The generated embedding for each application represents the corresponding API graphs where similar graphs have high cosine similarity. 
The scatter plots of the generated embedding for both the datasets, i.e., Debrin and CICMaldroid using the t-SNE algorithm, are depicted in Fig. 4. Benign data points and malware data points are color-coded as green and red respectively. The figures demonstrated that the generated embedding is non-linear, and the level of overlap between the two classes is more in the case of Drebin than in the case of CICMalDroid.

\begin{table*}[h]
\centering
\begin{minipage}{\textwidth}
\caption{Performance result (in \%)}\label{overall performance}
\begin{adjustwidth}{2.5cm}{0cm}
\begin{tabular}{|*{10}{c|} }
    \hline
Model& Features & \multicolumn{4}{c|}{\textbf{CICMaldroid}} & \multicolumn{4}{c|}{\textbf{Drebin}}\\
\hline
  &   & Accuracy  &  Precision  &   Recall  &   F1-score  &  Accuracy  &  Precision  &   Recall &   F1-score \\
\hline
\hline
GCN & CF   & 94.27  &  96.08  &  96.45  &  96.26  &  92.75  & 58.65 & 69.18 & 63.48  \\
 GS  & CF  & 95.50  & 96.91  & 97.231  &  97.07  &  97.58 & 94.31 & 79.91 & 86.52   \\
NB & PI  & 95.53  & 96.26  & 97.96  &  97.1  &  92.54 & 60.99  & 66.66 & 63.71  \\
DT & PI  & 96.47  & 98.03  & 97.34  &  97.68  &  97.52 & 88.52  & 85.96 & 87.22   \\
RF & PI  & 97.53  & 98.39  & 98.39  &  98.39  & 98.12 & 94.32 & 86.08  & 90.00   \\
SVM & PI  & 97.10 & 98.17  & 98.04  &  98.1  &  97.84 & 93.83 & 83.49 & 88.36  \\
CNN      & PI  & 96.45  & 97.93  & 97.43  &  97.68  &  97.81 & 88.90  & 87.67 & 88.28 \\
GS+NB & PI+GE  & 94.28  & 99.13  & 93.33  &  96.1  &  93.21  &  60.36  & 90.00 & 72.25  \\
GS+DT & PI+GE  & 97.24  & 98.24  & 98.15  &  98.2  &  97.64  &  86.76  & 89.63  & 88.17 \\
GS+RF & PI+GE  & 96.97 & 98.21  & 97.82  &  98.02  & 98.57 & 95.45 & 89.75 & 92.51   \\
GS+SVM & PI+GE  & 97.93  & 98.73  & 98.55  &  98.64  & 98.65  & 95.54 & 90.48 & 92.94 \\
GS+CNN & PI+GE  & \textbf{98.33}  & \textbf{99.18}  & \textbf{98.60} &  \textbf{98.89}  & \textbf{98.68} & \textbf{95.27} & \textbf{91.08} & \textbf{93.13}  \\
\hline
\end{tabular}
\end{adjustwidth}
\footnotetext{PI: Permission and Intent, CF: Centrality Feature of API, GE: Graph Embedding; GCN: Graph Convolutional Neural Network, GS: Graph Sage, NB: Naive Bayes, DT: Decision Tree, RF: Random Forrest, CNN: Convolutional NN, SVM: Support Vector Machine}
\end{minipage}
\end{table*}

\begin{figure}
\centering
\includegraphics[width=9cm,]{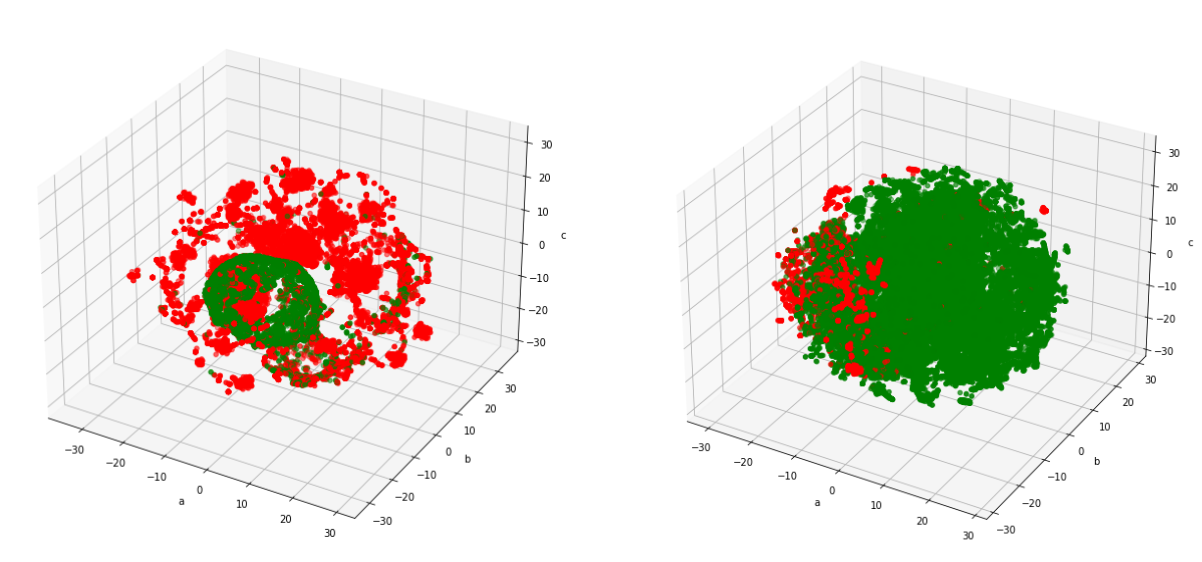}
\caption{\small{Scatter plots of the embedding generated by GraphSAGE from CICMaldroid (left) and Drebin Dataset (right).}}
\label{before_re-training}
\end{figure}

\subsubsection{Malware Detector} 
\begin{figure} [!hbt]
\centering
\includegraphics[width=8cm]{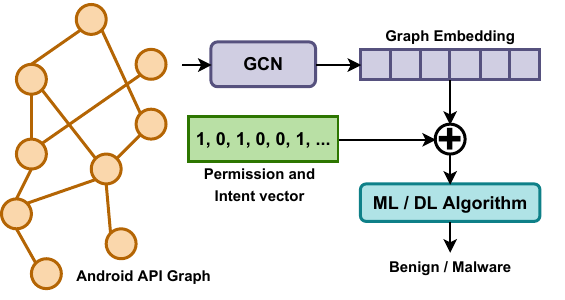}
\caption{Overall architecture of Malware classifier}
\label{Overall_archi}
\end{figure}

To detect malicious applications, we experimented with different types of machine learning algorithms combined with the graph embedding generated by the Graph Neural Network algorithm. The overall architecture of the malware detector is shown in Fig. \ref{Overall_archi}. Each Android application is first converted into an Android API graph using the graph construction technique described in Section \ref{preliminary_study_1}-B. The five centrality features of each node extracted in section \ref{preliminary_study_1}-B act as the feature of each node. The API graph is then converted into a graph embedding using GCN or GraphSAGE. The GNN variants, namely GCN and GraphSAGE, are trained separately in a supervised manner using the label of each application to generate the graph embedding. The generated graph embedding concatenates with Permission and Intent features. The combined features trained multiple machine learning algorithms, namely, Naïve Bayes, Decision Tree, Random Forest, SVM, and 1D-CNN. We also experimented with individual machine learning and deep learning algorithm using either the centrality feature or Permission and Intent as features to train the algorithm. The unique experiments are done to compare the results.

\paragraph{Naive Bayes (NB)}
The features are divided into two parts namely Permission and Intent, and graph embedding. First, using Permission and Intent we estimate the probability of benign and malware for each application using a Bernoulli Naive Bayes classifier. Second, graph embedding is then used as a feature to a Multinomial Naive Bayes classifier to estimate each application's probability of being benign and malware. The probability from both the classifier is used to train a final Multinomial Naive Bayes classifier to get the final output. 

\paragraph{Decision Tree (DT)}
DT is a machine learning algorithm that approximates a discrete-value target function. The Decision tree uses permission, intents, and graph embedding to construct the target function represented as a tree.
\paragraph{Random Forest (RF)}
It is a form of bagging technique where multiple decision trees are used. The multiple decision trees are trained on a random sampling of training observations, and uses random subsets of features for splitting nodes. In this work, we use 50 decision trees. The final predictions are made by averaging the predictions of multiple decision trees.

\paragraph{Support Vector Machine (SVM)} It is a machine learning algorithm that uses a decision boundary for classification. 

In this work, we use radial basis function kernel and the value of parameters C and Gamma are set to 10 and 0.1 respectively. The algorithm uses permission, intent, and graph embedding as features for training.

\paragraph{Convolutional Neural Network (CNN)} 
 In this work, we use global max pooling for flattening the output. Global max pooling is similar to max pooling, but the size of the window \say{f}  is equal to the length of the input. To train the CNN one-hot encoded vector, Permissions and Intents of an application is first to pass through an embedding layer to create an embedding that converts high dimensional data of Permissions and Intents into lower-dimensional vector space. The output of the embedding layer is sent to a dropout layer to prevent overfitting. The dropout layer's output is passed to the convolutional layers. There exists a total of six convolutional layers with kernel size set to 128, filter size set to 5, and padding set to SAME. Relu is used as an activation function, and the dilation rate is set to 1. Each convolutional layer is followed by a 1d max-pooling layer of pool size 2, except for the last convolutional layer, which is followed by global max pooling. The output of the global max-pooling layer is concatenated with the graph embedding generated using GraphSAGE Network. Finally, the concatenated input is passed through a fully connected layer to obtain the output of classification. In this experiment, we used Adam with default values to optimize the parameters of the network and categorical cross-entropy to calculate the loss of the network.

\section{Proposed Method}\label{proposed_method}
The GNN method developed in section \ref{preliminary_study_1} acts as a classifier with no gradient access to the attacker. To attack this model, we developed VGAE-MalGAN that can generate adversarial Android API graphs. 

\subsection{Preliminaries}

\paragraph{Variational Auto Encoder}
 Variational Auto Encoder (VAE) is a latent variable model. The model assumes that the observed data $A$  is generated from an unobserved latent variable $Z$. The aim is to capture intrinsic patterns in the observed data. The model can be thought of as describing the underlying process of generating A from the latent variable $Z$ using a probability distribution $p(A|Z)$. An ideal model will assign a high probability to observed $A$. Assuming $p(A|Z)$ is parameterized by $\theta$ it needs to solve the following optimization problem
 
 \begin{equation}
     \max_{\theta}^{}p_{\theta}(A)
 \end{equation}
 
 where, $p_{\theta}(A)=\int_{z}{}p(Z)p_{\theta}(A|Z)$, but this is an intractable integral over $Z$. To solve this problem the problem is instead formulated as inferring posterior $p(Z|A)$.This too involve an integral over $Z$, since $p(Z|A)=\frac{p(A,Z)}{\int_{z}^{}p(A,Z)}$. However, this can be solved using variational inference, which converts the problem into an optimization problem of finding an approximate probability $q(Z|A)$ close to $p(Z|A)$. VAE approximate the posterior probability $q_{\phi}(Z|A)$ and $p_{\theta}(A,Z)$ using a neural network, where $\phi$ and $\theta$ are the parameters of the network. It also assumes that the approximate posterior is a multivariate Gaussian $\mathcal{N}$ with a diagonal covariance matrix known as the re-parametrization trick. The parameters of this Multilayer Perceptron are calculated using the neural network with two nonlinear functions $\mu_{\phi}$ and $\sigma_{\phi}$. The following equation can formalize this.
 
 \begin{equation}
     q_{\phi}(Z|A)=\mathcal{N}(Z;\mu_{(\phi)}(A),\sigma_{\phi}(A)I)
 \end{equation}
 
 where, $I$ is the Identity Matrix.
 
 For the generator $p_{\theta}(A,Z)$, it is assumed $p(Z)$ is a fixed unit multivariate Gaussian i.e $p(Z)=\mathcal{N}(0,I)$ and $p_{\theta}(A|Z)$ is given by the following equation.
 
 \begin{equation}
     p_{\theta}(A|Z)=\mathcal{N}(A;\mu_{(\theta)}(Z),\sigma_{\theta}(Z)I)
 \end{equation}
 
 The summary of VAE network architecture is shown in equation 8:
 
 \begin{equation}
     A\xrightarrow[]{q_{\phi}(Z|A)}Z\xrightarrow[]{p_{\theta}(A|Z)}A
 \end{equation}
 
\paragraph{Variational Graph Auto Encoder}
Variational Graph Auto Encoder (VGAE) \cite{kipf2016variational} is a version of VAE that uses GraphSage to estimate $q_{\phi}(Z|A)$ and $p_{\theta}(A, Z)$ instead of a neural network since the neural network cannot work with Graph-based input. In this work, a modified version of VGAE is used. In the original proposal, the decoder part of the VGAE is susceptible to generating densely connected graphs. On the other hand, we found that the Android API graph generated in Section \ref{preliminary_study_1} is sparsely connected. Another reason for modifying the original VGAE model is to preserve the malware characteristic of the generated adversarial API graph with only valid nodes and edges that we have encountered in the dataset. Section \ref{proposed_method} gives more details on the implemented version of VGAE. 

\paragraph{Generative Adversarial Network}
Generative Adversarial Network (GAN) was initially proposed in \cite{goodfellow2014generative} where two networks play a min-max game. The first network, known as the generator, attempts to create data from the original distribution, whereas the second network, known as the discriminator, attempts to determine whether the data comes from the original distribution. The two networks train alternatively where the generator constantly tries to fool the discriminator, and the discriminator constantly detects fake data.

In this work, the generator attempts to generate adversarial malware samples that can evade the detection, whereas the discriminator updates itself to detect adversarial malware samples from the generator and data from the dataset. The whole training stops when the discriminator cannot identify adversarial samples generated by the generator.

\subsection{Threat Model}
The important components of an Android malware classifier that an attacker can have as background knowledge are as follows:
\begin{enumerate}
    \item Features: This component indicates whether the attacker knows about the type of features used for classification. In the case of Android, there are several features such as API, Permission, Strings of code, etc. The knowledge of the features can help manipulate and force the classifier to produce the wrong classification. 
    \item Model: This component indicates whether the attacker knows about the type of classifier used. The classifier can be any machine learning or deep learning algorithm which helps the attacker exploit its weakness.
    \item Weights/Parameter of the model: This component indicates whether the attacker knows about the weights (parameter $\theta$). The weights decide the decision boundary in classification, and having access to the weights can help the attacker craft the features to make the classifier produce the wrong classification.
    \item Dataset: This component indicates whether the attacker knows about the type of dataset used for training the classifier. Suppose an attacker knows the details of the dataset, the attacker can experiment and find out the result of the classifier and accordingly change the input to produce the wrong classification output. 
\end{enumerate}

\subsection{Problem Definition}
Let $A={A_{1}.....A_{n}}$ denote all the API graph adjacency matrices in the dataset. $\mathrm{A}_{n}^{mal}$ denotes the $n^{th}$ malware application whereas $\mathrm{A}_{n}^{ben}$ denotes the $n^{th}$ benign application. The detector is a mapping function $b:A \to\{ {0,1}\}$, where  $b(\mathrm{A}_{n}^{ben})=0$ and $b(\mathrm{A}_{n}^{mal})=1$. 

The substitute detector of VGAE-MalGAN aims to learn the function $s:A \to\{ {0,1}\}$, where  $s(\mathrm{A}_{n}^{ben})=b(\mathrm{A}_{n}^{ben})$ and $s(\mathrm{A}_{n}^{mal})=b(\mathrm{A}_{n}^{mal})$ whereas the generator learns to find a latent variable $Z$ that generates adversarial malware examples $\hat{A}$ where $s(\hat{A})=0$. Since the substitute detector aims to mimic the fuction $b$, the aim of the generator is to learn a latent variable $Z$ that generates  $\hat{A}$ where $s(\hat{A})=0$ and $b(\hat{A})=0$ with the constraint of preserving the malware functionalities of $\hat{A}$.

\subsection{Proposed VGAE-MAlGAN}
\begin{figure}
\centering
\includegraphics[width=8.7cm]{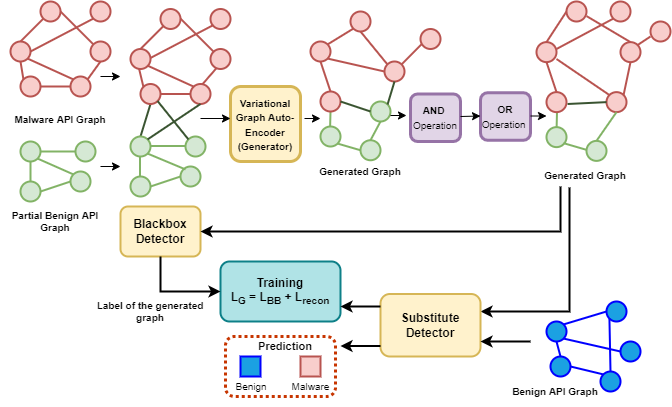}
\caption{{\small{Overall architecture of VGAE-MalGAN}}}
\label{VGAEMalGAN}
\end{figure}

 Recent years have shown the growth of adversarial malware examples generation using GAN dynamically. One of the main works is MalGAN proposed in \cite{hu2017generating}. MalGAN focuses on binary features where if the App calls $i^{th}$ API, the $i^{th}$ feature vector is set to 1; else it is set to 0. These binary features are fixed-sized in length with a fixed feature ordering. Models such as MalGAN are not usable for relational data such as  API graphs since the API graphs have no fixed size with no ordering of nodes. The proposed method named VGAE-MalGAN can effectively add nodes and edges to an existing  API graph to fool a detector based on GNN. Fig. \ref{VGAEMalGAN} shows the overall architecture of VGAE-MalGAN. The model considered is a GNN model, and the malware author has no access to the detector except for its prediction. The work assumes that the malware author knows that the detector is a GNN based model and knows the features used.
The proposed model consists of a generator and a substitute detector. The generator is a modified version of Variation Graph Auto Encoder, and the substitute detector is a GraphSAGE model. The substitute detector tries to fit the detector, and the goal of the Generator is to produce  API graphs that can fool the substitute detector. Since the substitute detector tries to fit the detector, fooling the substitute detector eventually leads to fooling the detector.
\subsubsection{Generator}
The Generator aims to produce API graphs that can fool the detector by inserting nodes and edges to an existing Android malware graph. The work assumes a latent Variable $Z$ exists underlying the process of benign data generation. The generator is a two-layer GraphSAGE parameterized by weights $\theta_{g}$. The generator takes as input the adjacency matrix of the original malware graph denoted by $A$ with randomly inserted nodes and edges generated from a randomly chosen benign example from the dataset. The goal of inserting randomly chosen nodes and edges from the benign dataset is to make the generator aware of the crucial nodes that influence the substitute detector in classifying an API Graph as benign or malware. This work employs a modified version of Variation Graph Auto Encoder (VGAE) as a generator. The encoder part of VGAE encodes a given  API graph into a latent Variable Z parameterized by the mean $\mu$ and standard deviation $\sigma$. The decoder then uses the latent variable $Z$ to reconstruct the graph adjacency matrix.
The encoder consists of two-layer GraphSAGE. The GraphSAGE layer generate a low dimensional representation of the Graph denoted by $Z$ given the adjacency matrix $A$ and feature matrix $X$ using the following equation.
\begin{multline}
    q(Z\|X,A)=\prod_{i=1}^{N}q(z_{i}\|X,A) \\ \ \  \textrm{with} \ \ q(z_{i}\|X,A)=\mathcal{N}(z_{i}\|\mu_{i},diag(\sigma_{i}^{2}))
\end{multline}

where, $\mu=GCN_{\mu}(X,A) = \tilde{A}\bar{X}W_{1}$, $log\sigma^{2}=GCN_{\sigma}(X,A) = \tilde{A}\bar{X}W_{1}$,  
$\bar{X}=GCN(X,A) = RELU(\tilde{A}XW_{0})$ and $\tilde{A}=D^{-\frac{1}{2}}AD^{-\frac{1}{2}}$ is the symmetrically normalized adjacency matrix.

The decoder takes the latent variable $Z$ and generates the adjacency matrix of the Graph $\hat{A}$. In this work, instead of using the decoder in the VGAE \cite{kipf2016variational} given by equation 10, we used the decoder given in equation 11. Our experimental analysis observed that the original decoder is very susceptible to densely connected Graph, which is not the case for the API graph. $Z$ is a $M*M$ square matrix where $M$ is the number of nodes. Entry in the generated adjacency matrix $\hat{A}_{i,j}$ is set to 1 if $e^{-dist_{L2}(z_{i},z_{j})}$ is greater than \say{0.98}, otherwise it is set to \say{0}, where $dist_{L2}(Z_{i},Z_{j})$ is the euclidean distance between \say{$z_{i}$} and \say{$z_{j}$}.

\begin{equation}
    \hat{A}=\sigma(ZZ^{T})
\end{equation}

\begin{equation} 
\hat{A}_{i,j} = 
\begin{cases}
  1 \ \ \text{if $\ \ e^{-dist_{L2}(Z_{i},Z_{j})}>0.98$}
  \\
  0 \ \ otherwise
\end{cases}
\end{equation}

After generating the adjacency matrix, we perform an AND operation with the Global adjacency matrix to ensure the decoder includes only valid edges. The AND operation keeps edges between API valid since the Global adjacency matrix \say{O} contains only valid edges from the entire dataset. After the AND operation, the resultant adjacency matrix performs OR operation between the resultant adjacency matrix and the original malware adjacency matrix. The OR operation makes sure the malware behavior is kept intact. The operation is shown in equation 12 and equation 13.

\begin{equation}
    \hat{A}=\hat{A} \ \ \& \ \ O
\end{equation}

\begin{equation}
    \hat{A}=\hat{A} \ \ \| \ \ A
\end{equation}

\subsubsection{Substitute Detector}
 The substitute detector trains to fit the detector so that the gradient can propagate to the Generator, which then can use the gradient to adjust the generator weights to generate API graphs that can fool the detector. Benign examples from the dataset and malware examples generated using the generator train the substitute detector. The probability of a given graph $G$ predicted as malware by the substitute detector $D$ is denoted as $D_{\theta_{d}}$.

\subsubsection{Training VGAE-MalGAN}
The training of VGAE-MalGAN is aimed to generate API graphs that can fool the detector. In order to train VGAE-MalGAN, the generator and the substitute detector trains simultaneously. Firstly, the dataset's malware samples are passed through the Generator to generate adversarial malware API graphs. The generated adversarial Android malware API graph from the dataset uses the detector to get the corresponding label for each generated adversarial Android malware API graph. The substitute detector uses the label given by the detector as ground truth to mimic the detector. For benign example, the substitute detector uses ground truth data from the dataset for training. Equation 14 gives the loss function of the substitute detector.
\begin{multline}
   L_{D}=-\mathbb{E}_{A\in Benign} log(1-D_{\theta_{d}}(A))\\ - \mathbb{E}_{\hat{A}\in BBMalware}  log(D_{\theta_{d}}(\hat{A}))  
\end{multline}

where, $D_{\theta_{d}}(.)$ is the predicted probability of the  API graph $\hat{A}$ as malware by the substitute detector, $\mathbb{E}_{A\in benign}$ is the expected value over all benign API graph and $\mathbb{E}_{\hat{A}\in BBMalware}$ is the expected value over all generated adversarial malware API graph detected as malware by substitute detector. 
The substitute detector minimized $L_{D}$ with respect to its weights to mimic the detector. Equation 15-17 gives the Loss of the generator denoted by $L_{G}$.
\begin{equation}
\small{
L_{BB}  =\mathbb{E}_{\hat{A}\in BBMalware} \log (D_{\theta_{d}}(\hat{A}))
}
\end{equation}
\begin{multline}
  L_{recon}=\mathbb{E}_{q(Z\|X,A)}[\log p_{\theta_{g}}(\hat{A}\|Z)] \\
  -  KL[q(Z\|X,\hat{A})\| p(Z)]  
\end{multline}

\begin{equation}
\small{
L_{G}=L_{BB}+L_{recon}
}
\end{equation}

where, $D_{\theta_{d}}(\hat{A})$ is the predicted probability of the generated adversarial malware adjacency matrix $\hat{A}$ as malware by the substitute detector, $\log p(\hat{A}\|Z)$ is the reconstruction loss of the generated adversarial malware adjacency matrix from the latent variable $Z$, $KL[q(.)\|p(.)]$ is the Kullback-Leibler divergence between $q(.)$ and $p(.)$ with a Gaussian Prior $P(Z) = \prod_{i}p(z_{i})=\prod_{i}\mathcal{N}(z_{i}\|0,I)$. Minimizing $L_{G}$ requires the $L_{BB}$ and $L_{recon}$ to be minimized. Since $L_{BB}$ represents the loss of the generated adversarial malware API graph as malware by the substitute detector and $L_{recon}$ is the encoder loss of VGAE. When both the losses are low, the encoder part of VGAE has effectively produced malware examples that can fool the Substitute detector into thinking the adversarial malware example as benign. Since the substitute detector trains to fit the detector, the generated API graph with its malware capabilities preserved eventually fools the detector. Algorithm 1, shows the training of VGAE-MalGAN. Once the training ends, the VGAE can sample adversarial Android malware API graphs using $p(\hat{A}\|Z)$.

\begin{algorithm}
    \caption{Training VGAE-MalGAN}
    \scriptsize{
    \begin{algorithmic}[1]
    \While {not converging}
    \For{API graph $A$ in malware train set}
     \State $B^{'}$=randomly select a benign API graph from benign train set
     \State noise=extract half of the edges and vertexes from $B^{'}$
     \State $A^{'}$=concatenate $B^{'}$ and noise
     \State $\hat{A}$= generate graph using VGAE and $A^{'}$ as input
     \State Label $\hat{A}$ using the detector 
    \EndFor
    \State Update the weights $\theta_{d}$ by descending along the gradient of $\bigtriangledown _{\theta_{d}}L_{D}$ using the benign API graph from the train set and generated API graphs $\hat{A}$ as input
    \For{API graph $A$ in malware train set}
     \State $B^{'}$=randomly select a benign API graph from benign train set
     \State noise=extract half of the edges and vertexes from $B^{'}$
     \State $A^{'}$=concatenate $B^{'}$ and noise
     \State $\hat{A}$= generate graph using VGAE and $A^{'}$ as input
     \State Update the weights $\theta_{g}$ by descending along the gradient $\bigtriangledown _{\theta_{g}}L_{G}$ using $\hat{A}$ as input
    \EndFor
    \EndWhile
     \end{algorithmic}
    } 
\end{algorithm}

\begin{figure}[h]
\centering
\includegraphics[width=9cm]{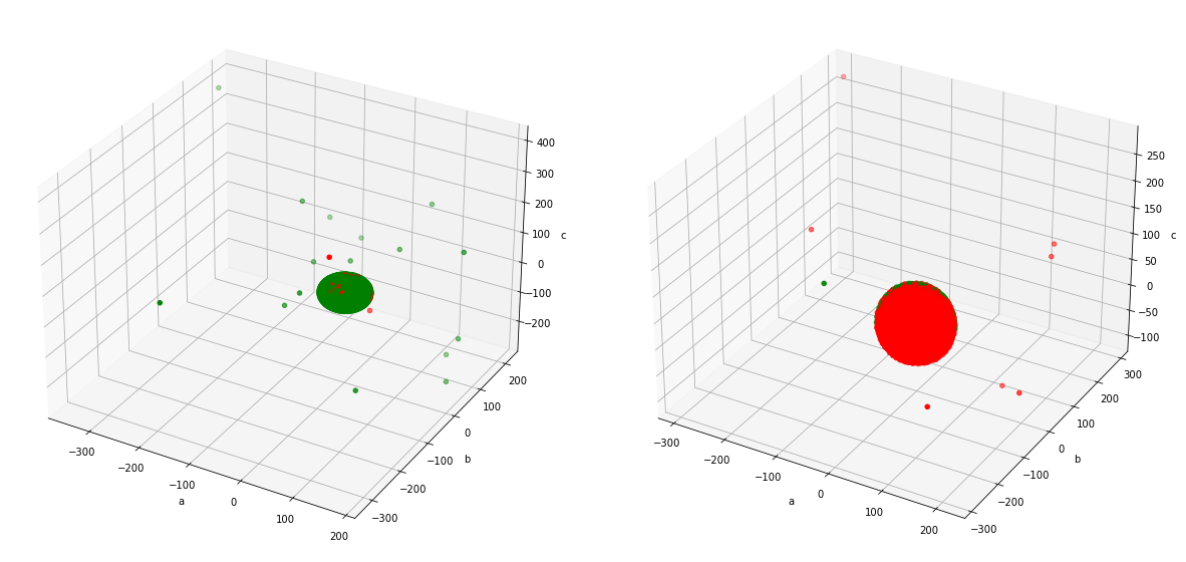}
\caption{\small{Scatter plots of the embedding generated from CICMaldroid dataset (left) and Drebin dataset (right) using substitute detector}}
\label{Sub_emb}
\end{figure}

\begin{figure}[h]
\centering
\includegraphics[width=9cm]{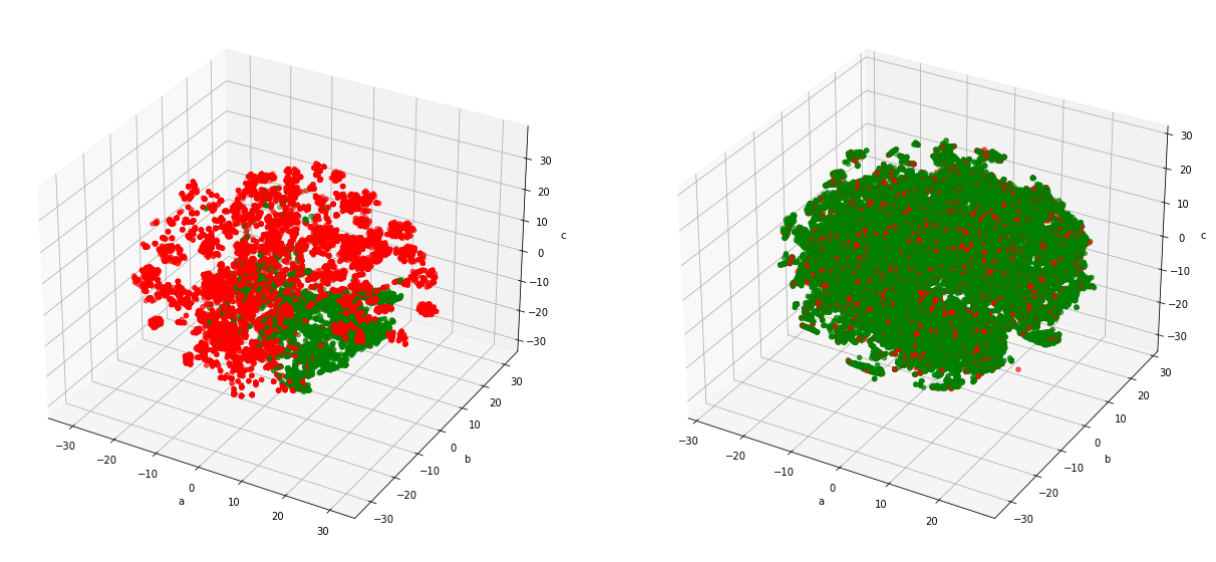}
\caption{\small{Scatter plots of the embedding generated from CICMaldroid dataset (left) and Drebin dataset (right) after retraining}}
\label{retain_emb}
\end{figure}

\subsection{Defense against VGAE-MalGAN}
To protect against manipulated API graph attack using VGAE-MalGAN, administrator can use VGAE-MalGAN to generate adversarial malware examples. The collected adversarial malware examples can then be labeled as malware and combined with existing dataset to retrain the model. Retraining helps the classifier to learn exploited relationship and thus harden the model against such attacks.

\section{Evaluation}\label{evaluation}
\subsection{Experimental Setup of VGAE-MalGAN}
\subsubsection{Dataset} The dataset used to test the effectiveness of VGAE-MalGAN is similar to the dataset we used in our preliminary study in Section \ref{preliminary_study_1} where it describes the details of the dataset.

\subsubsection{Experimental Result of Malware Detection}
In order to evaluate the performance of the deep learning model we choose Accuracy, Precision, and Recall.
Accuracy (A) is the ratio of the number of correct predictions to the total number of predictions.

Precision (P) is the ratio of correctly predicted positive observation to the total predicted positive observation or how many are actual malware out of what we predicted as malware.

Recall (R) is the ratio of correctly predicted positive observation to all observations in ground truth positive or how many did we predict correctly out of all the malware in the dataset.

F1-Score (F1) is the harmonic mean of precision and recall.

To understand the effect of using graph embedding generated using GNN, we first trained all the algorithms using only permission and intents. Table \ref{overall performance} summarises the overall result. From the experimental result shown in Table \ref{overall performance}, it is evident that when graph embedding was used along with Permission and Intent, most of the classifiers performed better than classifiers trained using only Permission and Intent. 

The relationships captured by constructing the API graph and extracting centrality features help generate graph embedding that encapsulates graphs' characteristics. The graph embedding help improve the representation generated using CNN for Android malware classification. CNN trained with Permission, Intent, and graph embedding achieves the best result among all the classifiers with an impressive F1-Score of 98.89\% and 93.13\% on CICMaldroid2020 and Drebin, respectively. The comparisons with other works is shown in Table \ref{performance comparision} and the proposed method is comparable to other state-of-the-art methods.

\subsubsection{Substitute Detector and its hyperparameters}
The substitute detector (SD) is a three-layer GraphSAGE network of size $(5 \times 32)$, and $(32 \times 32)$ parameterized by weights $\theta_{d}$. After the GraphSage layer, a global mean pool layer converts the output into a $1 \times 32$  vector, which is the graph embedding of the application. The Adam optimization algorithm was used for optimizing the network parameters with the learning rate set to 0.001, beta1 set to 0.9, beta2  set to 0.999, and epsilon set to 1e-08. The network's loss is calculated using the cross-entropy function and mean aggregation as the aggregator for GraphSAGE during training. In order to demonstrate that the substitute detector need not be similar to the attacked model, we experimented with a GCN network as the substitute detector to attack the GraphSAGE Model. The GCN network has two layers of size $(5 \times 32)$ and $(32 \times 32)$ followed by a global mean pool layer. The Adam optimization was used for optimizing the network parameters as before with the same parameter values. The default hyperparameters usually implemented in common deep learning libraries are the ones chosen.

\subsection{Experimental Result and Analysis}

Malware Adversarial attack is a severe threat to malware detection systems that use machine learning to differentiate malware and benign applications. Adversarial malware API graphs generated using VGAE-MalGAN could fool the malware detector under different background knowledge scenarios as in Table II. 


\subsubsection{Scenario 1}
In this scenario, the attacker knows the feature used in classification of the Android application's API graph and the APIs used to construct the API graph. 
The centrality features using the dataset are calculated since the attacker can access the dataset and the API graph. In terms of the model, the attacker knows the version of the GNN model used for classification. The trained GraphSAGE network in Section \ref{preliminary_study_1} was used as the model to be attacked. As shown in the malware detection performance in Table \ref{overall performance}, the GraphSAGE network was chosen as it gives higher detection than GCN. The substitute detector is a three-layer GraphSAGE network described in section \ref{evaluation} (A).
In another experiment, to demonstrate that the substitute detector need not be similar to the attacked model, we used a GCN network. The architecture of the substitute detector is explained in section \ref{evaluation} (A).
In this scenario, our attack is effective and can significantly reduce the malware detection rate of the attack model, as shown in Table \ref{VGAE performance}.

\subsubsection{Scenario 2} In this scenario, the attacker knows the feature used is the Android application's API graph and the APIs used to construct the API graph. In this scenario, we assume that the attacker only has access to 30\% of the original dataset. Using the partial dataset, we calculate the centrality features of the nodes in the API graph. Similar to scenario 1, the model to be attacked is GraphSAGE model described in section \ref{preliminary_study_1}. We experimented with  GraphSAGE as the substitute detector. The network architecture is similar to the one described in scenario 1. In this scenario, our attack is effective and can significantly reduce the malware detection rate of the attack model, as shown in Table \ref{VGAE performance}.

\subsubsection{Scenario 3}
In this scenario, the attacker knows that GNN is part of the classification process combined with another model. The attacker also knows that the features used is the Android application's API graph and the APIs used to construct the API graph. Based on the results shown in Table \ref{overall performance}, we choose the model with the highest performance in Android malware classification. Therefore, the model to be attacked combines GraphSAGE and CNN, and the model's details can be found in section \ref{preliminary_study_1}. Since the work focuses on attacking the GNN part of the model, we experimented with GraphSAGE as the substitute detector. The network architecture is similar to the one described in scenario 1. In this scenario, our attack is effective and can significantly reduce the malware detection rate of the attack model, as shown in Table \ref{VGAE performance}.

\begin{table}
\label{table-scenarios}
\centering
\scriptsize
\caption{\small{Background Knowledge to evaluate VGAE-MalGAN}} 
\begin{tabular}{lllll}
\hline\noalign{\smallskip} 
 Scenario&Feature&Model& Model Parameter & Dataset\\
\noalign{\smallskip}\hline\noalign{\smallskip}
1 &	Yes &	Yes &	No &	Yes	 \\
2 & Yes & Yes & No & Partial \\
3 & Yes & Partial & No & Yes\\

\noalign{\smallskip}\hline
\end{tabular}
\end{table}

\subsubsection{Evaluation Summary Under the Three Scenarios}
When the GraphSAGE network is used as a substitute detector to attack the GraphSAGE model in Scenario 1, it performs slightly better than using the GCN  network as a substitute detector. Since the embedding calculation of the graph is slightly different between GraphSAGE and GCN, the result is expected. However, the attack is still effective even when we attack the GraphSAGE classifier model using GCN as the substitute detector. From the experimental result shown in Table \ref{VGAE performance}, it is evident that when using only GraphSAGE as a substitute detector in the Drebin dataset, the recall (actual malware identification rate) is reduced by around 27\% for the Drebin dataset. For CICMaldroid, recall is reduced by around 80\%. When GCN replaces the substitute detector to attack the GraphSAGE model, we see that the model's recall is reduced by 24\% in the case of Drebin and 81.83\% in the case of CICMaldroid. From the results of the experiment in Scenario 2, shown in Table \ref{VGAE performance}, it is evident that when the attacker has access to a partial dataset which is assumed to be 30\% of the total dataset, the attack is still effective, although not as effective as having access to the entire dataset. In this experiment, the model's recall is reduced by 19.61\% in Drebin and  67.13\% in the case of CICMaldroid. The experiment in Scenario 3 shows that VGAE-MalGAN can still attack a combination of models when GNN is part of the classification. In this experiment, Drebin's recall is reduced by around 25\%, whereas around 51.38\% is reduced from CICMaldroid recall. From the experiments, we observed that it is easier to fool the CICMaldroid dataset than Drebin. One potential reason could be that Drebin has more apps in the dataset than CICMaldroid, making the classifier learn more complex decision surfaces, and making it harder to fool the classifier using VGAE-MalGAN. From the results shown in Table \ref{VGAE performance}, GraphSAGE with CNN is more challenging to fool than only using GNN algorithms. From the experimental results, we have seen that it is easier to fool the classifier when the attacker has access to the entire dataset. We retrain the classifier using the newly generated Adversarial API graph generated using VGAE in Scenario 1 and the original dataset. Table \ref{retraining_performance} shows the result of the retrained classifier. The retrained classifier achieved a similar F1 score with a difference of less than 1\% to 2\% as in table I except for Drebin classified with GraphSAGE. The retrained classifier thus is made more resilient to Adversarial examples generated using API manipulation.


The embedding of the substitute detector is shown in Fig. \ref{Sub_emb}. During VGAE-MalGAN training, the generator is optimized to deceive the substitute detector during training and the substitute detector is optimized to detect malicious and benign application. If the VGAE-GAN is trained successfully, the Substitute detector should not be able to distinguish between benign and malicious application which can be clearly seen in the embedding generated. The generated embedding after retraining is plotted in Fig. 9 using the t-SNE algorithm, where green represents benign data points, and red represents malware data points. The goal of training adversarial network is to craft malicious application that are very similar to benign applications with its malware characteristic preserved. The embedding generated by GraphSAGE shown in Fig. 4 have some separation between the benign and malware data points. After retraining, the embedding generated by GraphSAGE have no clear separation between the benign and malicious application as shown in Fig. 9. Although there is high overlap between the two embeddings, the retrained model almost maintain its classification accuracy with slight reduction in accuracy. Hence we can conclude that model must have learned much more complex decision boundaries as compared to before retraining.

\begin{table*}
\caption{\small{Performance comparison with other works (in \%)}}\label{performance comparision}
\scriptsize
\begin{adjustwidth}{2.5cm}{0cm}
\begin{tabular}{lllllll}
\hline\noalign{\smallskip}
Scheme & Dataset & Benign/Malware APKs & Accuracy  &  Precision  &   Recall  &   F1-score\\
  \noalign{\smallskip}\hline\noalign{\smallskip}
 John, T. et al.(2021)\cite{john2020graph} & Drebin  &  1410/720 & 92.30 &   91.50  &  93.30   &  92.30\\
Zhang et al.(2019)\cite{zhang2019efficient} & Drebin  & 5,900/5,600   &   96.00 &   90.07  &  \textbf{95.00}  &  \textbf{96.00} \\
Bai et al.(2020)\cite{bai2020famd} & Drebin  &5,900/5,560  &96.00  &   \textbf{97.00}  &   95.00  &  96.00   \\
Our work & Drebin  &  50,901/5,600  &  \textbf{98.68} & 95.27 & 91.08 & 93.13  \\
Mahdavifar et al. (2020)\cite{mahdavifar2020dynamic} & CICMaldroid  &  1479/11,598  &   96.70  &   99.16  &  96.54  &  97.84    \\

Alenezi et al. (2021) \cite{alenezi2021explainability} & CICMaldroid & 1479/11,598 & 94.70 & 93.00 & 94.00 & 93.00\\
Zhang, W (2021) \cite{zhang2021android} & CICMaldroid & 5687/5826 & 95.44 & 95.45 & 95.45 & 95.44 \\
Our work & CICMaldroid & 3,696/12,152 & \textbf{98.33}  &  \textbf{99.18}  &  \textbf{98.60}  & \textbf{98.89}\\
\noalign{\smallskip}\hline
\end{tabular}
\end{adjustwidth}
\footnotetext{\\}
\end{table*}

\begin{table}
\centering
\scriptsize
\caption{\small{Original recall and recall after training VGAE-MalGAN (in \%)}}\label{VGAE performance}
\begin{tabular}{llllll}
\hline\noalign{\smallskip}
 Scenario &Dataset &Classifier & Original & Attacked & SD\\
 &&&Recall&Recall&\\
\noalign{\smallskip}\hline\noalign{\smallskip}
1 &	Drebin &	GS &	79.91 &	52.5	&	GS \\
1 &	CICMaldroid	& GS &	97.23 &	11.33	&	GS \\
1 &	Drebin &	GS &	79.91 &	55.2	&	GCN \\
1 &	CICMaldroid	& GS &	97.23 &	15.4	&	GCN \\
2 &	Drebin &	GS &	79.91 &	60.3	&	GS \\
2 &	CICMaldroid	& GS &	97.23 &	30.1	&	GS \\
3 &	Drebin	& GS+CNN &	91.08 &	65.64	&	GS \\
3 &	CICMaldroid	& GS+CNN &	98.6 & 47.72 &	GS \\

\noalign{\smallskip}\hline
\end{tabular}
\footnotetext{GS: GraphSage, CNN: Convolutional NN \\}

\end{table}

\begin{table}
  \centering
\caption{\small{Performance after retraining using graph generated by VGAE-MalGAN (in \%)}}\label{retraining_performance}
\begin{tabular}{llllll}
\hline\noalign{\smallskip}
Dataset& Model & Accuracy & Precision & Recall & F1\\
\noalign{\smallskip}\hline\noalign{\smallskip}
Drebin & GS  & 96.47 & 90.66 & 70.96 &79.61  \\
Drebin & GS+CNN  &  98.43 & 92.92 & 91.01 &  91.96\\
CICMaldroid & GS  & 94.24 & 96.05 & 96.44 & 96.25 \\
CICMaldroid & GS+CNN & 97.86 & 98.76 & 98.46 & 98.61  \\
\noalign{\smallskip}\hline
\end{tabular}
\footnotetext{GS: GraphSage, CNN: Convolutional NN\\}
\end{table}

\subsection{Comparison with State-of-the-Art Attack}
Recently, many algorithms have been proposed to attack collective classification algorithms and Graph Neural networks. Recent works in \cite{xu2020attacking} and \cite{wang2019attacking} have proposed algorithms to attack target nodes in the context of collective classification of nodes in a graph by creating nodes and connecting to existing nodes or by adding or deleting edges. In the context of GCN nettack \cite{zugner2018adversarial}, it has been proposed to attack the target node by either modifying the graph structure or the node attributes. It is also shown in \cite{wang2019attacking} that GCN model-specific algorithm such as nettack is better in reducing the classification accuracy of the nodes than algorithms designed to attack collective classification. These algorithms focus on changing the target of a node in the graph. Algorithms that focus on node classification cannot be applied in our work since APIs represented as nodes do not have a target. All the APIs used are benign. The only difference between APIs used in benign and malicious applications is how the APIs are used collectively. In the context of graph classification, recent work in \cite{wan2021adversarial} proposed an algorithm called Grabnel. Grabnel is a Bayesian optimization-based attack method for graph classification models with three attack modes: creating/removing an edge, rewiring or swapping an edge, and node injection. Although this algorithm can successfully attack the model, it has no mechanism to preserve the original code as in our attack. In our model, we preserve the original graph by performing an AND operation with the global adjacency matrix only to add valid edges between nodes and an OR operation with the original Malware API, which makes sure that the original semantics of the malware API graph is preserved.

\section{Related Works}\label{related_works}
 Incorporating deep learning techniques to identify Android malware has recently become a popular research domain. In this section, we review the different feature representations and machine/deep learning techniques for identifying Android malware. The three main analysis types for Android malware identification are static analysis, dynamic analysis, and hybrid analysis. The most common features used for static analysis include Requested Permissions, Intent, API calls, and App components. In dynamic analysis, app actions, execution paths, and network features are the standard features used. Hybrid analysis used both static and dynamic features. In static analysis, the research works in \cite{su2016deep},\cite{su2020droiddeep} use Permissions, Sensitive API calls, Intent, and App components as features. In dynamic analysis, the research works in \cite{yuan2016droiddetector}, and \cite{yuan2014droid} use information like action\_sendnet, which sends data over the network, as features. In recent years, the graph-based approach has become popular to study relationships within the App. Instead of just studying sequential features, studying existing relationships within an application has proven to be helpful in identifying malware. As an example, MalScan \cite{wu2019malscan} extracts extract function call graph, Hindroid \cite{hou2017hindroid} extracts API relationships based on Code block and API-Invoke method. Existing work in fixed-size features such as GAN attack on black box detector in \cite{hu2017generating} generates adversarial samples based on the sequence of the binary vector of API calls by preserving the malware characteristics. A bi-objective GAN consisting of two discriminators, one to distinguish malicious examples and one to distinguish adversarial examples from normal, is proposed in \cite{li2019adversarial}. The work uses sequences of permissions, actions, and APIs of the Android application as a feature vector to train the GAN. Image-based classification and GAN-based attack are proposed in \cite{wang2021advandmal} where system calls of API are used as features to generate RGB images and uses pix2pix adversarial network to generate adversarial examples. Opcode-based image and a GAN is used to generate adversarial example in \cite{chen2021using}.
 

\section{Conclusion}\label{conclusion}
Chatzoglou et al. in their work \cite{Chat2022} examined more than forty IoT based Android official apps statically and dynamically and found that majority of the apps have a range of security and privacy issues. There were repeated incidents of malicious code injection into popular Android apps through advertising SDKs, as in the case of CamScanner in 2021 \cite{malreport}. The Android banking trojans are working with new capabilities. Ransomware such as FLocker is capable of locking Android TV sets \cite{IOTAndroid}. Another notable one was Android app known as Dresscode which steals data. When a device infected with Dresscode comes in contact with a network with weak router password, it can infect other devices including IoT devices \cite{IOTAndroid2}. So IoT based Android malware detection is becoming all the more important. In this work, graph embedding based on centrality measures is generated using GNN and combined with Permission and Intent to train multiple machine learning algorithms for Android malware classification. The API graph construction and centrality feature extraction help to generate effective graph embedding that encapsulates the difference between malware and benign applications. From the results, we concluded that graph embedding helps to improve the feature representation, thus increasing the overall performance. A new architecture named VGAE-MalGAN is introduced and experimented upon to show that it can effectively reduce the malware detection rate of GNN based classifiers. VGAE-MalGAN can generate adversarial samples that can help to combat such attacks through retraining the model. More analysis on effective graph embedding generation, incorporating different kinds of features that are resistant to adversarial attacks will be looked at in the future work.

%

\vspace{11pt}

\vspace{-33pt}
\begin{IEEEbiography}[{\includegraphics[width=1in,height=1.25in,clip,keepaspectratio]{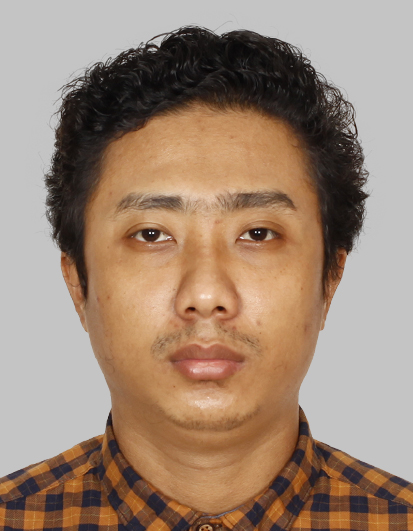}}]{Rahul Yumlembam} is currently pursuing PhD degree in Computer and Information Sciences at Northumbria University, UK. He finished BTech degree in Computer Science and Engineering and MTech degree in Computer Science and Engineering (Artificial Intelligence) in India. He was a Project Fellow at IIT, Guwahati, India. His research interest includes Machine Learning, Deep learning, Cyber Security using AI, Big Data, Brain-Computer Interface, Text Mining, and Image Processing.

\end{IEEEbiography}
\begin{IEEEbiography} [{\includegraphics[width=1in,height=1.25in,clip,keepaspectratio]{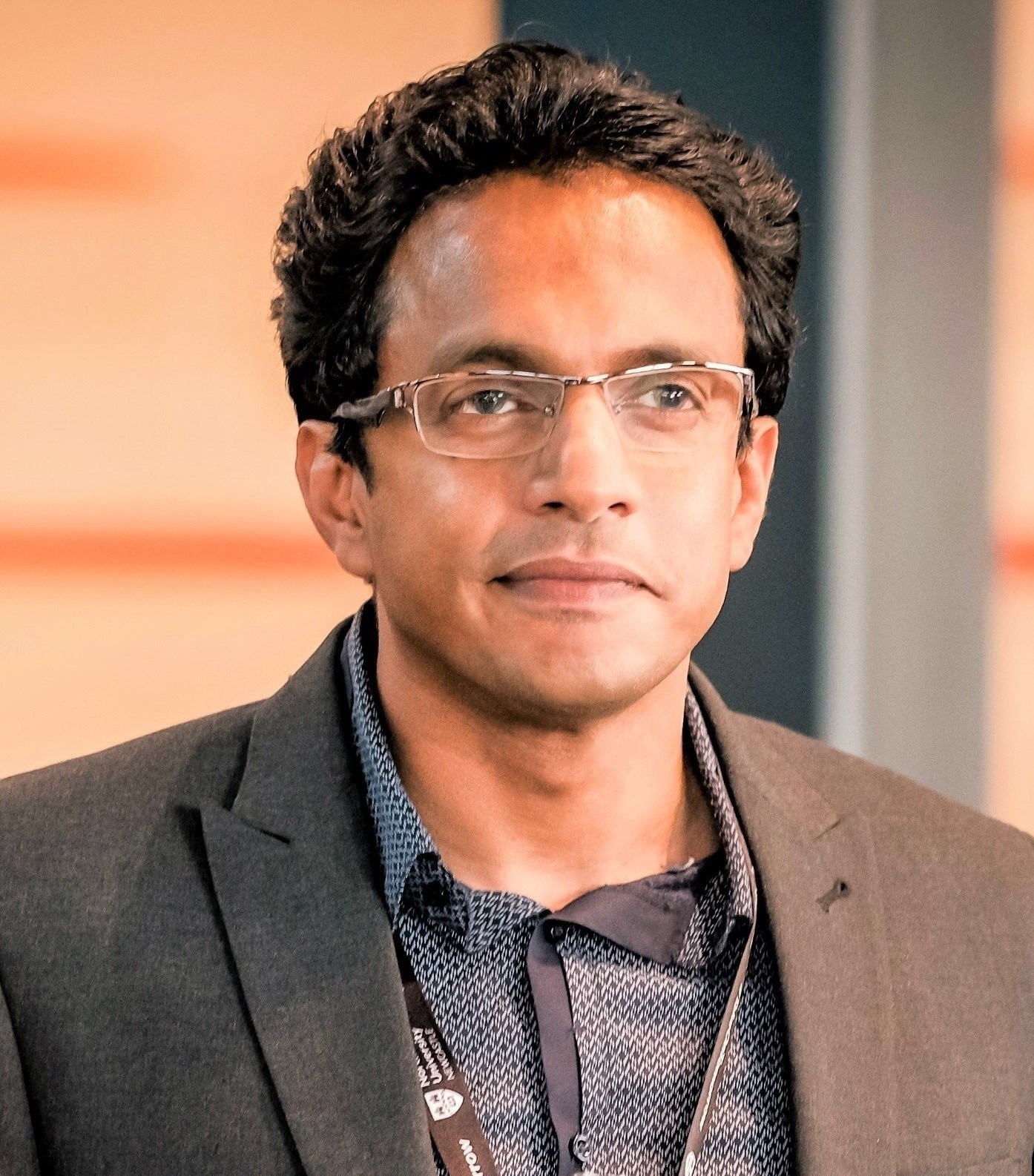}}]{Biju Issac} received the BE degree in Electronics and Communications Engineering, the Master of Computer Applications (MCA) degree, and the PhD degree in Networking and Mobile Communications. He is an Associate Professor in Northumbria University, UK. He is research active and has authored more than 100 refereed conference papers, journal articles, and book chapters. His research interests include Networks, Cybersecurity, Applied Machine Learning and Deep Learning.
\end{IEEEbiography}
\begin{IEEEbiography}[{\includegraphics[width=1in,height=1.25in,clip,keepaspectratio]{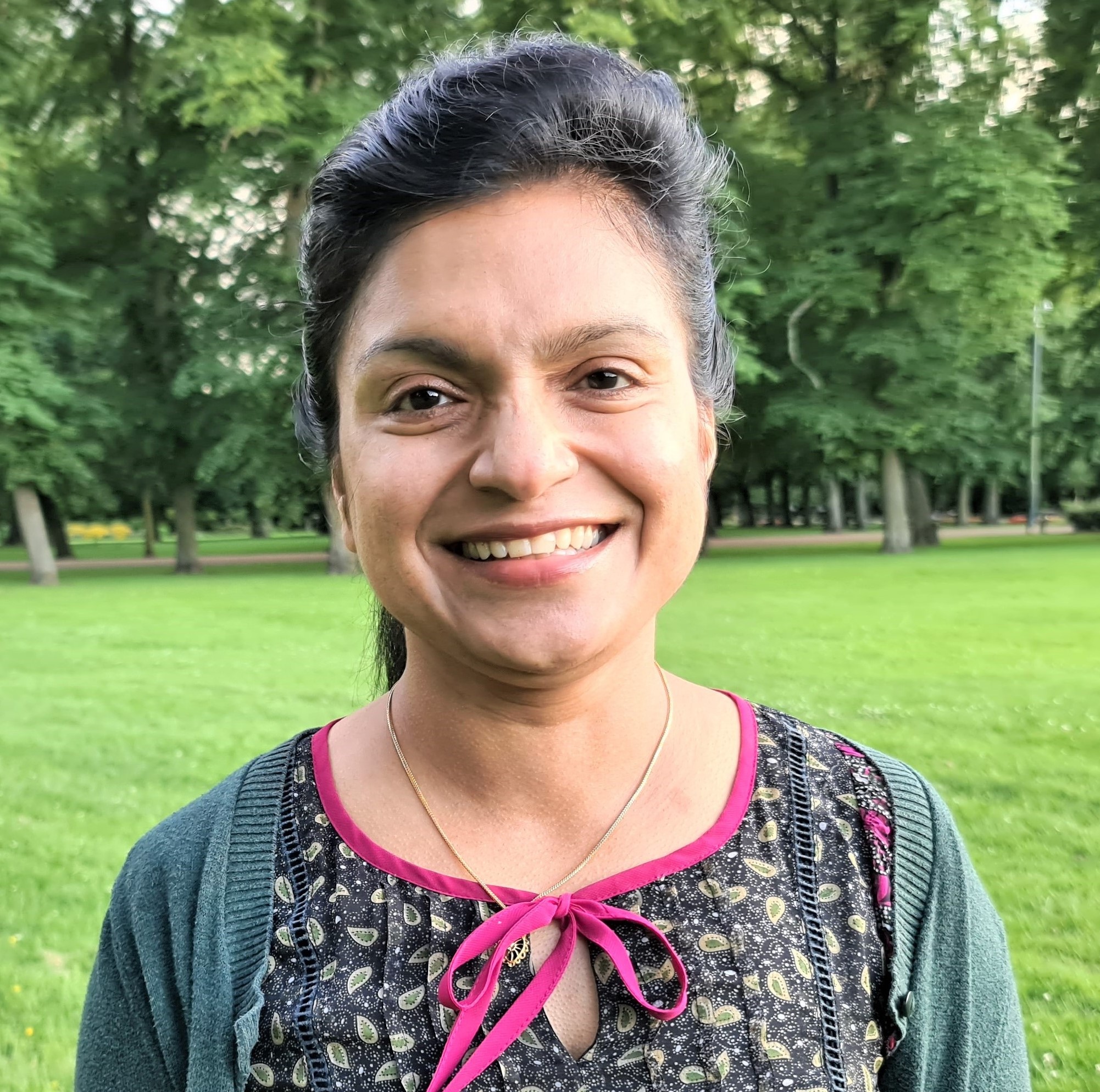}}]{Seibu Mary Jacob} received the BSc and MSc degrees in Mathematics, the Post Graduate Diploma in Computer Applications (PGDCA) degree, the Bachelor’s degree in Mathematics Education (BEd), and the PhD degree in Mathematics Education. She is currently a Senior Lecturer teaching Engineering Mathematics in Teesside University, UK. She has authored more than 30 research publications as book chapters, journal articles, and conference papers. 
\end{IEEEbiography}
\begin{IEEEbiography}[{\includegraphics[width=1in,height=1.25in,clip,keepaspectratio]{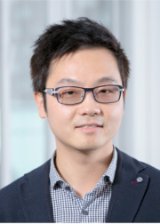}}]{Longzhi Yang} received the BSc degree in computer science from the Nanjing University of Science and Technology, China, in 2003, the MSc degree in computer science from Coventry University, Coventry, UK, in 2006, and the PhD degree in Computer Science from Aberystwyth University, Aberystwyth, UK, in 2011. He is a Professor in Northumbria University, UK. His research interests include computational intelligence, machine learning, big data, computer vision, intelligent control systems etc. 
\end{IEEEbiography}

\vfill
\vfill
\vfill
\end{document}